\documentstyle[epsfig]{aipproc}

\begin{document}

\ifx\href\undefined
\else\errmessage{don't use hypertex}
\fi

\title{The Evidence for Massive Star Formation in Early-Type Spiral Galaxies}

\author{Salman Hameed \& Nick Devereux}
\address{New Mexico State University\\
Las Cruces, New Mexico 88003-8001}
\maketitle

\begin{abstract}
A recent analysis of the IRAS database indicates that the massive star 
formation rates in early-type (Sa-Sab) spirals are comparable to the 
massive star formation rates in late-type spirals. $H\alpha$ imaging 
of some of the infrared luminous early-type spirals reveals two types of galaxies. One type shows clear signs of interaction, whereas the other 
type appears to host a nuclear starburst. The occurence of nuclear 
starbursts in early-type spirals may be related to the propensity for 
such galaxies to also host Seyfert nuclei. The evidence for interactions 
suggests that early-type spirals are evolving in the current epoch.
\end{abstract}

\section*{Introduction}
%
%
%
%
Early-type (Sa-Sab) spirals are widely perceived to be the most inert
of spiral galaxies. Indeed, the very term "early-type" refers to an early 
phase of star formation that has long since passed eg. \cite{sandage}. The perception is based, 
in large part, on the optical morphology which is dominated by an inert 
stellar bulge. There are other observational results, however, which 
suggest that early-type spirals are not as quiescent as once believed. One
of the most striking is the fact that early-type spirals are among the 
most luminous in the population of nearby galaxies (D $\le$ 40Mpc) when 
observed in the far-infrared with the Infrared Astronomical Satellite
\cite{devereux}. Adopting the far-infrared luminosity as a 
measure of the massive star formation rate, one finds that the IRAS data
reveals a previously unsuspected population of early-type spirals with 
star formation rates that rival the most prodigously star forming Sc 
galaxies.

\section*{The Hubble type dependence of the L(FIR)/L(Blue) ratio}

Data obtained with the Infrared Astronomical Satellite (IRAS) are used to 
investigate the dependence of the L(FIR)/L(Blue) ratio on Hubble type for a 
sample of 1462 galaxies selected from the Nearby Galaxy Catalog \cite{tully}.
Adopting the far infrared luminosity as a measure of the present day massive 
star formation rate and the blue luminosity as a measure of the past star formation 
rate, the L(FIR)/L(Blue) luminosity ratio measures the ratio of present to 
past star formation in a way that is analogous to $H\alpha$ equivalent widths. Our results (Fig. 1) \cite{devereux} indicate\\
~ ~ a) The mean L(FIR)/L(Blue) luminosity ratios are similar for Sa's, Sb's, 
and Sc's.\\
~ ~ b) The range in L(FIR)/L(Blue) luminosity ratios is as large for the 
early-type spirals as it is for the late type spirals.

Thus, L(FIR)/L(Blue) luminosity ratios do not support what may be one of the 
most basic tenets of extragalactic astronomy, that the ratio of present 
to past star formation is correlated with Hubble type.


\section*{$H\alpha$ imaging of luminous early-type spiral galaxies}

There is growing evidence to suggest that early-type spiral galaxies 
are perhaps the most dynamic of all the nearby galaxy systems. The discovery of ripples, shells \cite{schwiet}, and counter rotating gas and star disks in some early-type 
spirals \cite{braun} \cite{merr} indicate that a major accretion event occured in the past 1-2 Gyr \cite{hern}. 
We have obtained $H\alpha$ images for 5 of the most luminous ($L_{FIR} > 10^{10}L_{\sun}$) Sa-Sab galaxies. The $H\alpha$ images (Fig. 2) suggest two types of 
early-type spiral galaxies. One type clearly shows signs of disturbance in both the 
continuum and $H\alpha$ images. The other type includes galaxies 
that appear to be undisturbed morphologically, but their entire $H\alpha$ 
emission is radiated from a very small, $\le 1kpc$, diameter region centered 
on the nucleus. The origin of the nuclear $H\alpha$ emission is unclear, but 
it is most likely a starburst, an active nucleus or a combination of the two.

\section*{Summary}

The current perception that high mass star formation increases along 
the Hubble sequence (Sa-Scd) is largely based on measurements of the $H\alpha$
luminosities and equivalent widths by Kennicut \& Kent \cite{kenn}. Their 
measurements do indicate an increase in the ratio of present to past star formation along the sequence Sa to Sc. Our results, however, indicate 
that both the massive star formation rate and the present to past star 
formation ratio are   
independent of Hubble type, at least 
for the Sa's, the Sb's and the Sc's. On the other hand, there is still a lingering 
controversy over the origin of far infrared luminosity. We are therefore
obtaining complimentary $H\alpha$ images for a  
complete, all sky, sample of 57, bright, $m(B) \le 12$ magnitudes, nearby,
$ D\le 40 $ Mpc, early type (Sa-Sab) spiral galaxies in order to better 
understand the difference between the IRAS results and the existing 
$H\alpha$ measurements.

\begin{figure}[b!] 
\centerline{\epsfig{file=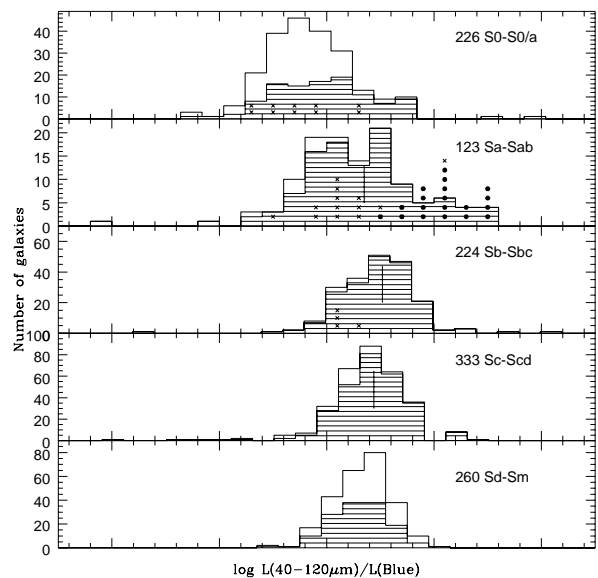,height=3.35in,width=3.35in}}
\vspace{10pt}

\caption{Histograms illustrating the Hubble type dependence of
$L_{fir}/L_{blue}$ratios. The hatched histograms identify FIR detections, the unshaded histograms identify upper limits to the FIR flux. The vertical bar identifies
the median of the distribution. Solid circles identify Sa-Sab galaxies with 
$L(40-120\mu m) \ge 10^{10} L_{\sun}$, and crosses identify Sa-Sab galaxies from Kennicutt's sample. Fourteen of Kenicutt's galaxies are classified as 
either S0-S0/a or Sb-Sbc in the NBG catalog. For details, see [2]}
\label{fig1}
\end{figure}

\begin{figure}[b!] 
\centerline{\hbox{\epsfig{file=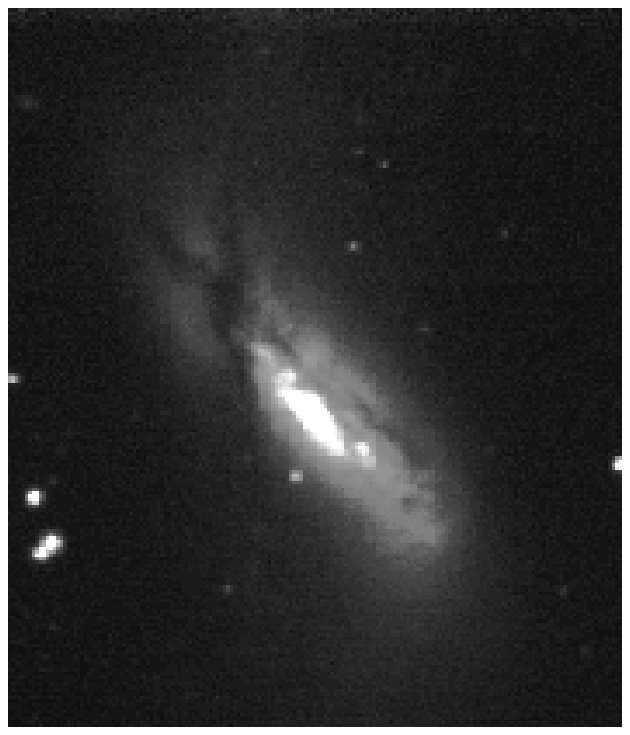}
\hspace {15pt}
\epsfig{file=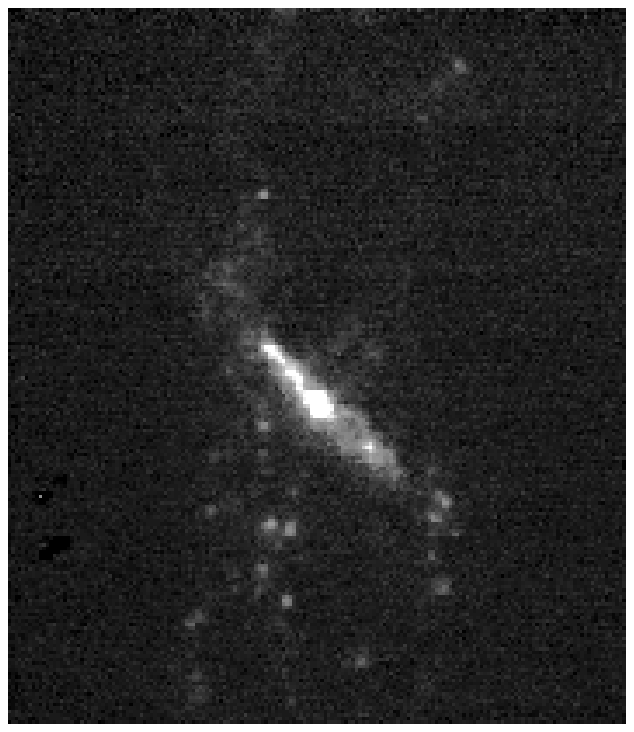}
}}
\vspace {15pt}
\centerline{\hbox{
\epsfig{file=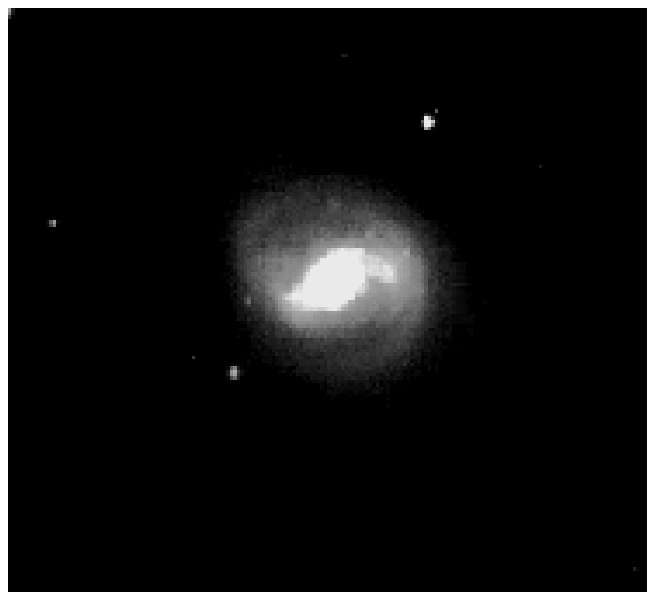}
\hspace {15pt}
\epsfig{file=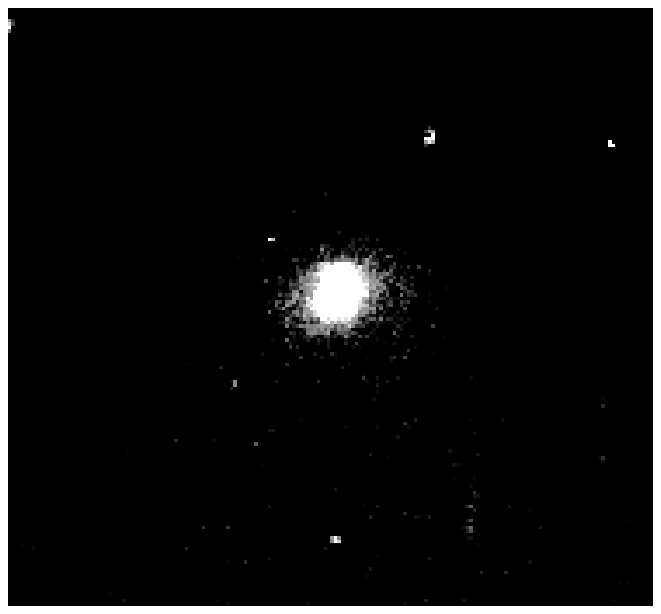}
}}
\vspace{10pt}
\caption{The continuum (left) and continuum subtracted $H\alpha$ (right) images of NGC 660 and NGC 1022 illustrate the two types of early-type spirals. 
NGC 660 shows clear signs of interaction, whereas NGC 1022 appears to host a nuclear 
starburst.}
\label{fig2}
\end{figure}


\begin{references}

\bibitem{sandage}Sandage, A. 1986, {\it A \& A.}\ {\bf
161}, 89.

\bibitem{devereux}Devereux, N.A., \& Hameed, S.A., 1997, {\it AJ.}\ {\bf
113}, in press.

\bibitem{tully}Tully, R.B., Nearby Galaxies Catalog, 1989 (Cambridge:Cambridge Univ. Press)

\bibitem{schwiet}Schweizer, F., \& Seitzer, P., 1988, {\it ApJ.}\ {\bf
328}, 88.

\bibitem{braun}Braun, R., et al. 1992, {\it Nature}\ {\bf
360}, 442.

\bibitem{merr}Merrifield, M.R. \& Kuijken, K., 1994, {\it ApJ.}\ {\bf
432}, 575.

\bibitem{hern}Hernquist, L. \& Spergel, D.N., 1992, {\it ApJ.}\ {\bf
399}, L117.

\bibitem{kenn}Kennicutt, R.C., Jr. \& Kent S.M., 1983, {\it AJ.}\ {\bf
88}, 1094.

\end{references}
\end{document}